# Large-area epitaxial growth of MoSe$_2$ via an incandescent molybdenum source

Man-Kit Cheng[1,#], Jing Liang[1,#], Ying-Hoi Lai[1], Liang-Xi Pang[1], Yi Liu[1], Junying Shen[1], Jianqiang Hou[1], Qing Lin He[1], Bochao Xu[2], Junshu Chen[2], Gan Wang[2], Chang Liu[2], Rolf Lortz[1], and Iam-Keong Sou[1,3,*]

[1] Department of Physics, the Hong Kong University of Science and Technology, Clear Water Bay, Hong Kong SAR, China

[2] Department of Physics, Southern University of Science and Technology, No. 1088, Xueyuan Rd., Xili, Nanshan District, Shenzhen, Guangdong, China

[3] William Mong Institute of Nano Science and Technology, the Hong Kong University of Science and Technology, Clear Water Bay, Hong Kong SAR, China

* To whom correspondence should be addressed. E-mail: phiksou@ust.hk

[#] Equal contribution

## Abstract

We have developed an incandescent Mo source to fabricate large-area single-crystalline MoSe$_2$ thin films. The as-grown MoSe$_2$ thin films were characterized using transmission electron microscopy, energy dispersive X-ray spectroscopy, atomic force microscopy, Raman spectroscopy, photoluminescence, reflection high energy electron diffraction (RHEED) and angular resolved photoemission spectroscopy (ARPES). A new Raman characteristic peak at 1591 cm$^{-1}$ was identified. Results from Raman spectroscopy, photoluminescence, RHEED and ARPES studies consistently reveal that large-area single crystalline mono-layer of MoSe$_2$ could be achieved by this technique. This technique enjoys several advantages over conventional approaches and could be extended to the growth of other two-dimensional layered materials containing a low-vapor-pressure element.

## Introduction

Recently two-dimensional (2D) layered materials are gaining increasing attention as they are highly promising for use in next-generation nanoelectronic and optoelectronic devices, which is attributed to their various intrinsic physical properties, and the fact that it is relatively easy to fabricate complex nanostructures from them. Because of its rich physics and high mobility, graphene is the most widely studied 2D material. However, the lack of a bandgap limits the application of pristine graphene in many of its promising applications, such as logic circuits, transistors and photonic devices. More recently, 2D layers of transition metal dichalcogenides (TMDCs), such as $MoS_2$, $WS_2$, $MoSe_2$, and $WSe_2$, have attracted much attention since they are promising to be the channel materials for field-effect transistors because of their bandgap comparable with silicon, structural stability, absence of dangling bonds, lack of short-channel effects and the ability to offer lower power consumption than in silicon-based classical transistors [1]. Being the selenide analogues of $MoS_2$ and $WS_2$, $MoSe_2$ and $WSe_2$ have smaller band gaps and higher electron mobilities, making them more appropriate for practical devices among these 2D layered materials.

Monolayer $MoSe_2$, in fact, is composed of three atomic layers, with a layer of Mo atoms sandwiched by two layers of Se atoms, and in multilayer structure, the sandwich-like units are bonded by weak Van der Waals interactions between Se atoms. $MoSe_2$ is a semiconductor with an indirect band-gap of 1.1eV in bulk and a direct one of 1.55eV in monolayers and bilayers. Its tunable interlayer coupling that might help enable bandgap tunability offers a good prospect for solar cell and optoelectronics applications.

To the best of our knowledge, so far fabrication of $MoSe_2$ thin films was limited to a few techniques such as micromechanical exfoliation [2,3], chemical vapor deposition [4-6] and molecular beam epitaxy (MBE) [7,8]. Due to the extremely low vapor pressure of Mo, MBE growth of $MoSe_2$ thin films usually requires the use of the e-beam evaporation technique, which requires expensive equipment and its growth uniformity is usually not desirable because of the limiting size of the heating area in the source. Currently, large-area fabrication of single-crystalline $MoSe_2$ and other TMDCs is still challenging. This will hamper further development of a number of large scale applications based on mono-layer TMDCs materials such as electricity generation [9] and sensing applications [10]. In this work we report the successful growth of large-area single-crystalline $MoSe_2$ thin films using an incandescent Mo source. Various characterizations of their structural, optical properties and electronic band structure were performed. In particularly the results from reflection high energy electron diffraction (RHEED), angular resolved photoemission spectroscopy (ARPES), Raman spectroscopy and photoluminescence (PL) demonstrate that large-area single crystalline mono-layer of $MoSe_2$ could be realized using this novel growth technique.

**Methods**

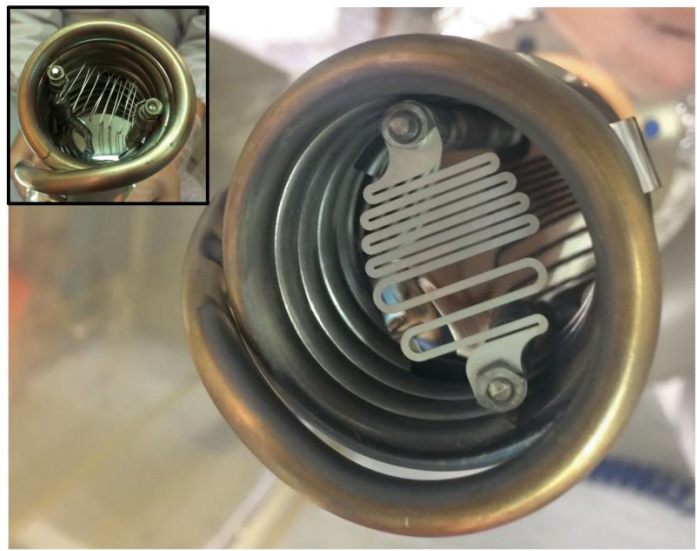

Figure 1: A homemade incandescent Mo source with a Mo-thin-foil filament. The inset shows the photo of a Mo-wire filament installed in the source.

The growth of MoSe$_2$ was carried out in an ultra-high vacuum (UHV) chamber previously used to store effusion cells. Unlike the standard MBE growth system, this modified UHV system does not equip with neither a liquid-nitrogen shroud nor a reflection high energy electron diffraction facility. A HeatWave Lab Model 101479-01 substrate holder capable of heating a 1 inch substrate up to 1200 ℃ is installed. A homemade incandescent Mo source and a standard Se effusion cell were used to handle the growth of MoSe$_2$ thin films. The incandescent Mo source, as shown in Figure 1, is based on Joule heating of a Mo filament cut from a high-purity Mo thin foil of 50 microns in thickness. Such a filament-type source owns several advantages over other growth approaches including 1) it could offer scalable large-area growth with high uniformity as the filament can be made to fit any desired size; 2) it offers high growth rate attributed to the large evaporation surface area of the source filament; 3) as a free-standing source it is contamination-free; 4) it is much less expensive. It is also worth mentioning that for the growth of 2D materials using such an incandescent source, the consumption of the filament is not a great concern since the required thickness for each growth is either a single layer or up to a few layers. If a retractable design for the incandescent source cell is used, filament replacing can be done without venting the growth chamber. Figure 1 shows a photo of the homemade incandescent Mo source, in which one can see that two tungsten electrodes are holding a Mo thin-foil filament and they are surrounded by a water cooling coil for radiation shielding. We have also used commercially available Mo wire as a replacement of the Mo thin-foil filament, the inset in Figure 1 shows the photo of a Mo-wire filament installed in the source. We have developed a formula that gives the temperature of the Mo filament versus the applied direct current, which assumes that the Joule heat power generates from filament is equal to its thermal radiation power and takes into account the temperature dependence of the resistivity [11] and emissivity [12] of the filament. All the MoSe$_2$ growths carried out in this study used an optimized condition for the Mo source with the filament current ranging from 7.45 A to 22.50 A and the filament temperature ranged from 1800 to 2100 °C as deduced from this formula, while the Se cell temperature was set at about 160 °C. Growth of MoSe$_2$ was conducted on several different substrates with the aim to optimize the surface morphology of the resulting MoSe$_2$ layer. These substrates include bare GaAs(111)B, GaAs(111)B with a Bi$_2$Te$_3$ buffer, Highly Ordered Pyrolytic Graphite (HOPG) and Sapphire substrates. The growth parameters used for fabricating the samples presented in this study are listed in Table 1. The thicknesses of Sample #1 - #5 were determined through their cross-sectional transmission electron microscopy (TEM) images while that of Sample #A was obtained by X-ray reflectivity. The thickness of Sample #B was estimated from the growth condition of Sample #A. The GaAs substrates used for the growth of Sample #1 - #4 were first preheated to about 580 °C for a few minutes aiming at removing their passive oxide layers. Preheating prior to the growth of MoSe$_2$ was also carried out for the HOPG and Sapphire substrates at 800 °C for 30 minutes and 600 °C for 30 minutes, respectively. The Bi$_2$Te$_3$ buffer layer of Sample #5 was grown after removing the passivation oxide layer of the substrate in a VG V80H MBE system followed by the growth of a Se capping layer, which was removed by preheating to 150 °C prior to the growth of the MoSe$_2$ thin film. The size of all the substrates used is about one-quarter of a two-inch wafer.

Table 1: Growth parameters used for samples fabrication

| Samples | Substrate | Substrate Temperature (°C) | Filament type | Filament Current (A) | Se cell Temperature (°C) | Growth Time (min) | MoSe$_2$ Thickness (ML) |
|---|---|---|---|---|---|---|---|
| 1 | n+ GaAs(111)B | 400 | 50 μm foil | 8.53 | 160 | 5 | 13 |
| 2 | GaAs(111)B | 420 | 300μm wire | 7.50 | 170 | 10 | 10-11 |
| 3 | GaAs(111)B | 420 | 300μm wire | 7.50 | 165 | 7.5 | 5-6 |
| 4 | GaAs(111)B | 415 | 50 μm foil | 8.50 | 160 | 1 | 1-2 |
| 5 | Bi$_2$Te$_3$/GaAs(111)B | 395 | 50 μm foil | 7.45 | 160 | 15 | 3 |
| 6 | HOPG | 535 | 50 μm foil | 8.20 | 160 | 3 | / |
| 7 | Al$_2$O$_3$(001) | 270 | 300μm foil | 18.5 | 165 | 60 | / |
| A | Al$_2$O$_3$(001) | 270 | 300μm foil | 22.5 | 170 | 8 | 6-7 |
| B | Al$_2$O$_3$(001) | 265 | 300μm foil | 18.5 | 160 | 40 | 1 |

The as-grown MoSe$_2$ samples were cut into small pieces for conducting various characterizations. Structural characterization was carried out using a JEOL JEM 2010F TEM with energy dispersive X-ray spectroscopy (EDS), and a Digital Instrument Nanoscope IIIa atomic force microscopy (AFM). Raman spectroscopic analysis and photo-luminescence were performed using a Renishaw Invia Micro Raman System. The ARPES measurements were performed using a laboratory-based ARPES system consisting of a SPECS PHOIBOS 150 electron analyzer and a UVLS-600 UV lamp at a pressure lower than $5 \times 10^{-10}$ mbar. The samples were cleaved in the

ARPES chamber, yielding a clean, flat surface. The incident photon energy was 21.218 eV (He I), the spot diameter was about 500 micrometers. Samples were found to be stable during a typical measurement period of ~ 20 hours.

**Results and Discussion**

Figure 2 describes the structural and chemical analysis of the $MoSe_2$ layers grown on GaAs(111)B substrates. Figure 2(a) displays the cross-sectional TEM images of a $MoSe_2$ sample (Sample #3) with 5-6 layers grown on a bare GaAs(111)B substrate. Figure 2(b) displays the cross-sectional TEM image of a $MoSe_2$ sample with 4 layers grown on a $Bi_2Te_3$/GaAs(111)B substrate (Sample #5), showing locally a smoother $MoSe_2$ thin film is realized attributed to the high smoothness of the $Bi_2Te_3$ buffer. Figure 2(c) displays the high resolution TEM image of zoomed in interface of $MoSe_2$ and GaAs of Sample #3 with its zone axis along the $[11\bar{2}0]$ direction of $MoSe_2$. As shown the lateral and vertical distance of $MoSe_2$ are 2.8 Å and 6.5Å, respectively. Corresponding schematic crystal structures of $MoSe_2$ drawn based on the measured distances in Figure 2(c) along the $[11\bar{2}0]$ (up) and [0001] (down) direction are shown in Figure 2 (d), revealing that the lattice parameters of the as-grown $MoSe_2$ layer should be a=3.23Å, and c= 13.0Å, in good agreement with the standard values of $MoSe_2$. Figure 2(e) displays the typical EDS profile obtained from a $MoSe_2$ layer with the built-in EDS spectroscopy in the high resolution TEM system, confirming the as-grown layer consists of Mo and much richer Se contents. The apparent Se/Mo ratio is 2.8 and not exactly 2. However, it is well known that EDS is a semi-quantitative technique and particularly for EDS performed in a cross-sectional TEM imaging mode of an ultra-thin layer, it becomes even less accurate. The well matched lattice parameters together with the results from Raman and photoluminescence studies to be addressed later offer strong evidence that the as-grown layer is in fact $MoSe_2$.

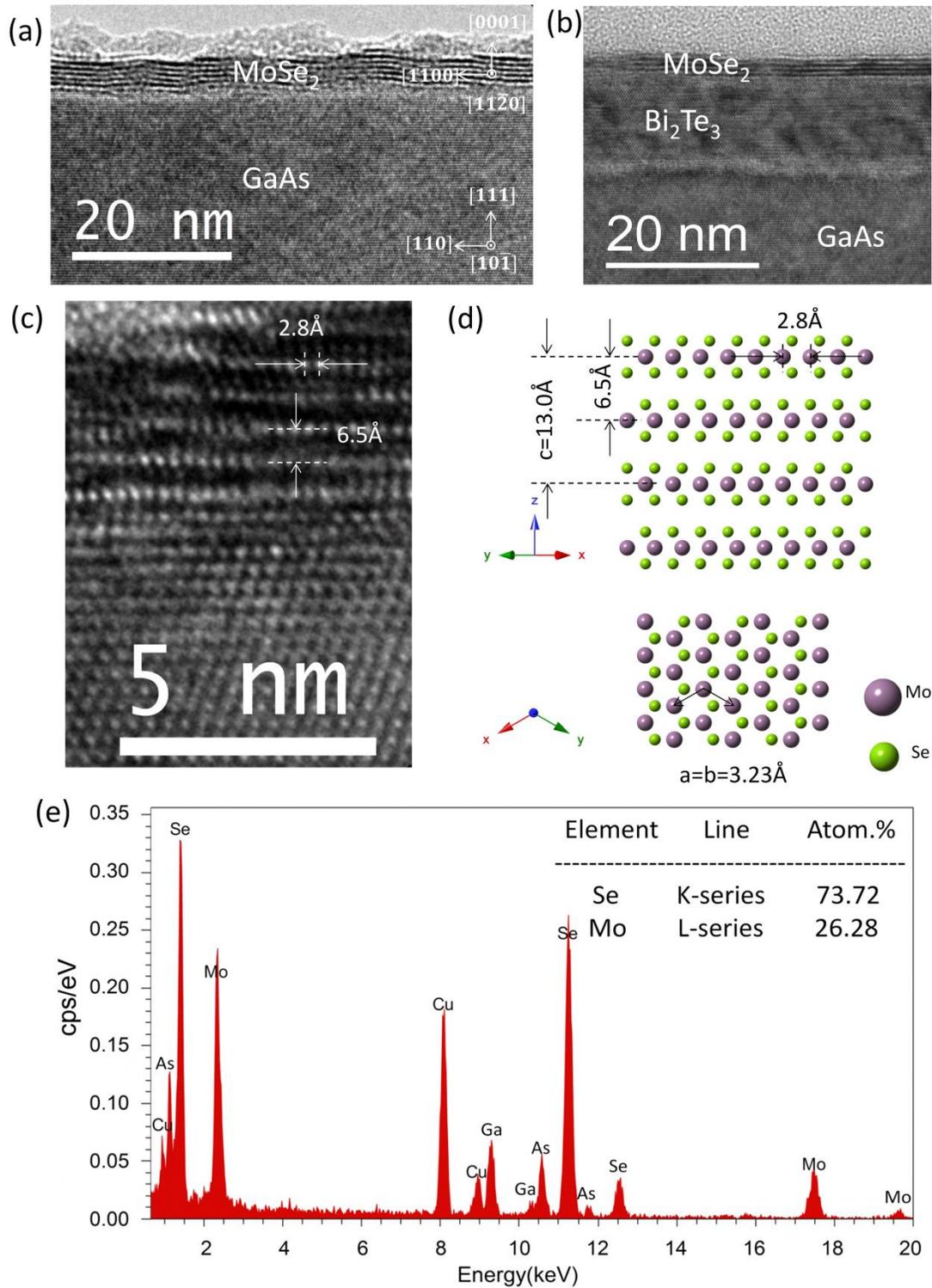

Fig. 2. (a) Cross-sectional TEM image of 5-6 ML MoSe$_2$ on a bare GaAs(111)B (Sample #3). (b) Cross-sectional TEM image of 4 ML MoSe$_2$ grown on a Bi2Te3/GaAs(111)B substrate (Sample #5), which results in a more smoother MoSe$_2$ layer. (c) High resolution TEM image of zoomed in interface of MoSe$_2$ and GaAs of Sample #3 with its zone axis along

the [11$\bar{2}$0] direction of MoSe$_2$ (d) Corresponding schematic crystal structure of MoSe$_2$ drawn based on the measured distances in (c) along the [11$\bar{2}$0] (up) and [0001] (down) direction. (e) Typical EDS spectrum taken for an as-grown MoSe$_2$ layer, where the Cu signal comes from the sample holder ring.

In order to reveal the large-scale surface morphology of the as-grown MoSe$_2$ samples, AFM surface profiling was conducted. Figure 3(a) and (b) show the typical AFM images with a scan area of $10 \times 10$ μm$^2$ and their corresponding line profiles for MoSe$_2$ layers grown on HOPG (Sample #6) and sapphire substrate (Sample #7), respectively (Figure S1(a) and (b) in supplementary information display the corresponding results for the MoSe$_2$ layers grown on a bare GaAs(111)B substrate and on a Bi$_2$Te$_3$ buffer). The corresponding AFM images and line profiles of a pure HOPG substrate and a pure sapphire substrate are shown in Figure S1(c) and (d) in supplementary information. Through a comparison among these AFM images, the surface morphology of the MoSe$_2$ layer grown on a HOPG substrate as shown in Figure 3(a) follows the main wavy surface profile of the substrate with an overall base surface roughness about a few nanometers. Islands with heights around 10 nm and density around 1-2 /μm$^2$ are also present. As shown in Figure 3(b), the best surface morphology is obtained for the MoSe$_2$ layer grown on a sapphire substrate. The AFM image of a pure sapphire substrate as shown in Figure S1(d), reveals some scratched lines resulted from the manufacturing process. The surface morphology of the MoSe$_2$ thin film grown on a sapphire substrate, as shown in Figure 3(b), basically is atomically flat except showing the scratched lines with a depth around one nanometer originated from the substrate plus just a few islands across the entire scan area of $10 \times 10$ μm$^2$.

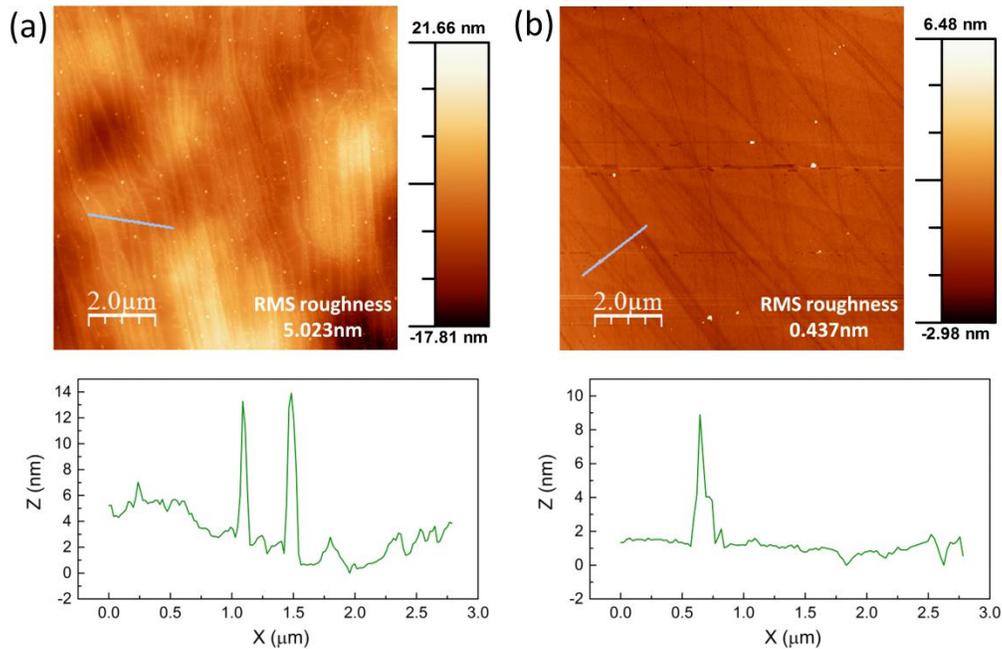

Figure 3: AFM images with $10 \times 10$ μm$^2$ scan area and their corresponding line profiles of MoSe$_2$ layers grown on (a) HOPG and (b) sapphire substrate.

Figure 4(a) displays the Raman spectroscopy spectra with the Raman shift ranging from 100 to 400 cm$^{-1}$ for Samples #1-#4 with different number of MoSe$_2$ layers grown on GaAs(111)B substrates using either Mo-thin-foil or wire filaments. As can be seen, the spectra of the MoSe$_2$ samples show less-intense LO and TO peaks of GaAs as the thickness of the MoSe$_2$ layer increases and three characteristic peaks at around ~170cm$^{-1}$, ~240 cm$^{-1}$ and ~350 cm$^{-1}$ of multilayer MoSe$_2$ are present. The inset in Figure 4(a) displays the characteristic peak of MoSe$_2$ around 240 cm$^{-1}$ for Sample #1 (13 ML), Sample #2 (10-11 ML), Sample #3 (5-6 ML) and Sample #4 (1-2 ML), in which one can see that the peak for Sample #4 with 1-2 ML of MoSe$_2$ is 1.4 cm$^{-1}$ lower than those of the other three samples with layer thickness over 5 ML. All the observations described above are in good agreement with the reported Raman shift values of a previous study [13].

Figure 4(b) shows the Raman spectroscopy spectrum for the MoSe$_2$ thin film grown on the Bi$_2$Te$_3$/GaAs(111)B substrate, in which the expected characteristic A$_{1g}$ peak of MoSe$_2$ together with the characteristic peaks of GaAs and Bi$_2$Te$_3$ are also easily identified. The origin of the peak around 170 cm$^{-1}$ has not yet been identified but it is believed to come from a BiTeSe alloy thin layer at the interface between the Bi$_2$Te$_3$ buffer and the MoSe$_2$ layer [14]. A wider spectral range covering Raman shifts from 100 to 1700 cm$^{-1}$ was then used to study the spectra of the MoSe$_2$ thin films grown on HOPG and sapphire substrates and the resulted spectra are displayed as the top curves in Figure 4(c) and (d), respectively. The two bottom curves in these figures are the spectra for a pure HOPG substrate and a pure sapphire substrate, respectively. As can be seen in Figure 4(c), five characteristic peaks of MoSe$_2$ grown on the HOPG substrate can be easily identified when compared with the spectrum of the pure HOPG substrate. All these five peaks are well matched with a previous report [15]. In Figure 4(d), one can also locate these peaks in the top curve for the MoSe$_2$ thin film grown on the sapphire substrate though several characteristic peaks of sapphire as shown in the bottom curve have a slight overlap with them. Interestingly, as the spectrum of the pure sapphire substrate in the spectral range between 1450 and 1700 cm$^{-1}$ is featureless, a new characteristic Raman peak of MoSe$_2$ is identified at ~1591 cm$^{-1}$, which has not been reported previously to the best of our knowledge.

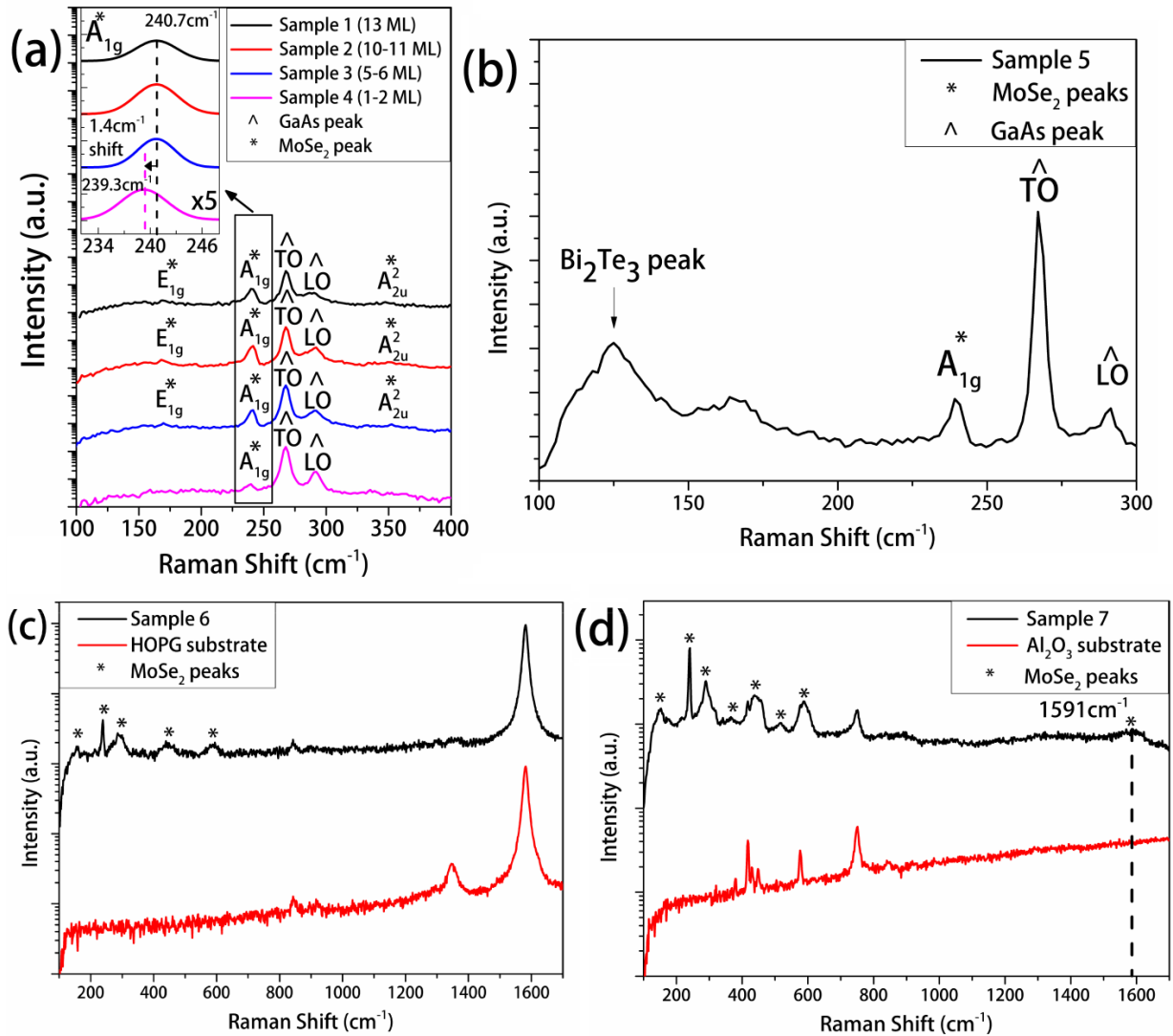

Figure 4: Raman spectra of MoSe$_2$ samples (a) Samples #1 – #4 with the inset shows the spectral region near the A$_{1g}$ peak at a finer scale; (b) Sample #5; (c) Sample #6 and (d) Sample #7.

Two MoSe$_2$ samples grown on sapphire substrates together with a pure sapphire substrate were studied by Raman spectroscopy and photoluminescence in details. The MoSe$_2$ layer of Sample #A has a thickness of 6-7 ML as determined by X-ray reflectivity (see Figure S2 in supplementary information). The other sample, Sample #B was grown using the condition listed in Table 1, based on the developed formula mentioned above, the growth of MoSe$_2$ with this condition is expected to have a thickness about 1 monolayer. Figure 5 shows the Raman Spectra of the two MoSe$_2$ samples as well as that of the pure sapphire substrate between 220 and 320 cm$^{-1}$. As can be seen the two peaks in this spectral range for Sample #A are located at 242.0 and 286.2 cm$^{-1}$ while those for Sample #B are located at 239.9 and 290.0 cm$^{-1}$. These data indicate that the peak at lower shift value of Sample # B is 2.1 cm$^{-1}$ lower than that of Sample #A while the peak at higher shift value of Sample #B is 3.8 cm$^{-1}$ higher than that of Sample #A. These observations are quantitatively matched with the results reported by P. Tonndorf *et al* [13] in which the Raman shift of a mechanically exfoliated monolayer MoSe$_2$ was compared with multiple-layers and bulk samples of MoSe$_2$, providing evidence that our Sample #B is likely as thin as one monolayer of MoSe$_2$. Another evidence of this claim comes from their resulted photoluminescence (PL) spectra as shown in Figure 6, in which one can see that only Sample #B gives a PL emission peak at around 783 nm while Sample #A does not offer any emission in this spectral range. P. Tonndorf *et al* [13] have also reported PL studies on MoSe$_2$ layers and the resulted PL peaks were found to be 792 nm and 807 nm for one and two layers exfoliated MoSe$_2$. As our observed PL peak of Sample #B is located at a wavelength a few nanometers shorter than 792 nm, plus the fact that its full-width half maximum is about 14 nm that is very close to that of the mono-layer sample in Ref. [13] (the latter is estimated to be 13 nm from the curve shown in Figure 6(a) in this reference), which further support that Sample #B likely contains only one monolayer of MoSe$_2$.

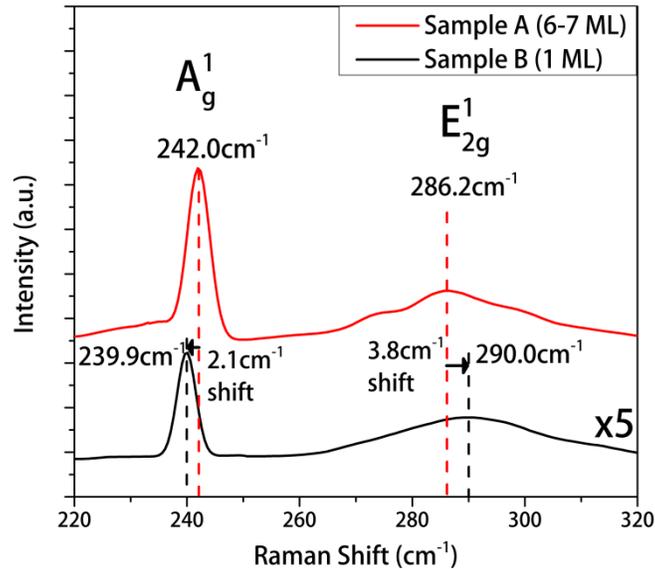

Figure 5: Raman Spectra of the two MoSe2 samples with thickness of 6-7ML(Sample #A) and 1ML(Sample #B).

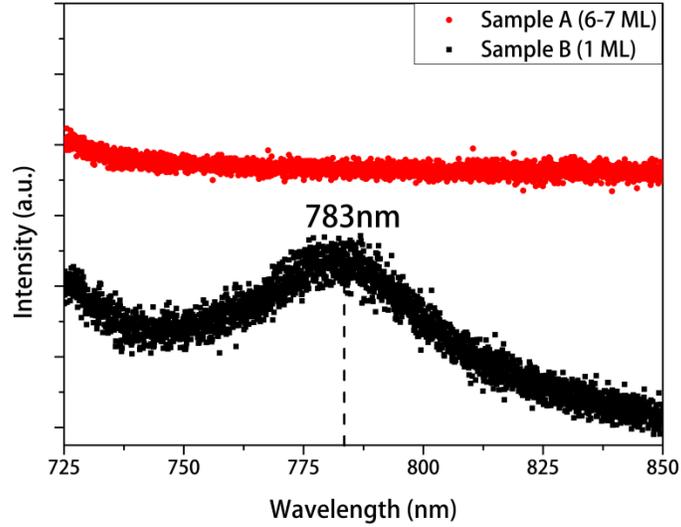

Figure 6: Photoluminescence (PL) spectra of MoSe2 with 6-7ML (Sample #A) and 1ML(Sample #B)

A more convincing confirmation of the mono-layer nature of the $MoSe_2$ layer of Sample #B comes from RHEED and ARPES studies. Sample #B was actually capped with a thin Se layer on top as a protection layer (Both the Raman and PL spectra of Sample #B were carried out after the Se cap has been removed by thermal heating in ultra-high vacuum). After growth, a piece of Sample #B with a size about 25 $mm^2$ was placed in the MBE system housed in the Department of Physics of the Southern University of Science and Technology. It was heated to 210 ℃ for 30 minutes with an e-beam heater to remove the Se capping layer while the reflection high energy electron diffraction (RHEED) facility was used as the monitoring tool. After preheating is done, streaky $MoSe_2$ RHEED patterns were seen everywhere within the sample, revealing that the whole piece being studied is single-crystalline. A typical RHEED pattern of the exposed $MoSe_2$ layer is shown in Figure S3 in the supplementary information. The sample was then transferred *in-situ* to the ARPES module of which the diameter of the light spot is 0.88 mm. Figure 7(a) displays the resulting ARPES spectrum and all the corresponding energy distribution curves along the Γ-K direction are plotted in Figure 7(b). Figure 7(c) displays the two energy distribution curves obtained at Γ and K points, showing the apices of valence bands at the Γ and K points with energy values of -2.3 eV and -2.09 eV from the fermi level, respectively. These two energy values are marked in Figure 7(a) by the two dash lines, indicating the apex at Γ point is 0.21 eV lower than the apex at K point. Previously, Y. Zhang *et al* [16] have reported a detail ARPES study on $MoSe_2$ epitaxial thin films grown on graphene with the aim to investigate the indirect-to-direct-bandgap the transition versus the thin film thickness. They reported that the apex at Γ point is lower than that at K point only occurred for $MoSe_2$ thin film with thickness smaller than 1 monolayer. Their observed energy difference between these two apices is 0.38 eV for one monolayer coverage of $MoSe_2$, which is larger than our observation. However, this difference could be attributed to the difference in the energy of light source, which is 70 eV in Y.

Zhang *et al*'s work [16] and 21.21 eV in our work. Our observed energy difference is in fact in good agreement with a more recent study by E. Xenogiannopoulou *et al* [17], who used the same light source as the one we used, in which the corresponding energy difference observed is in the range of 0.15 to 0.20 eV, for a one monolayer thick MoSe$_2$. In addition, as shown in Y. Zhang *et al*'s work [16], for a MoSe$_2$ thin film with thickness equal to or thicker than 1.2 monolayers, a narrow band split off from the valence band near the Γ point at energy of 1.17 eV below the fermi level should be detected. The fact that such a split off band is absent in our ARPES spectrum, further confirming that our MoSe$_2$ sample with the resulted ARPES spectrum shown in Figure 7(a) is likely with a thickness close to one monolayer.

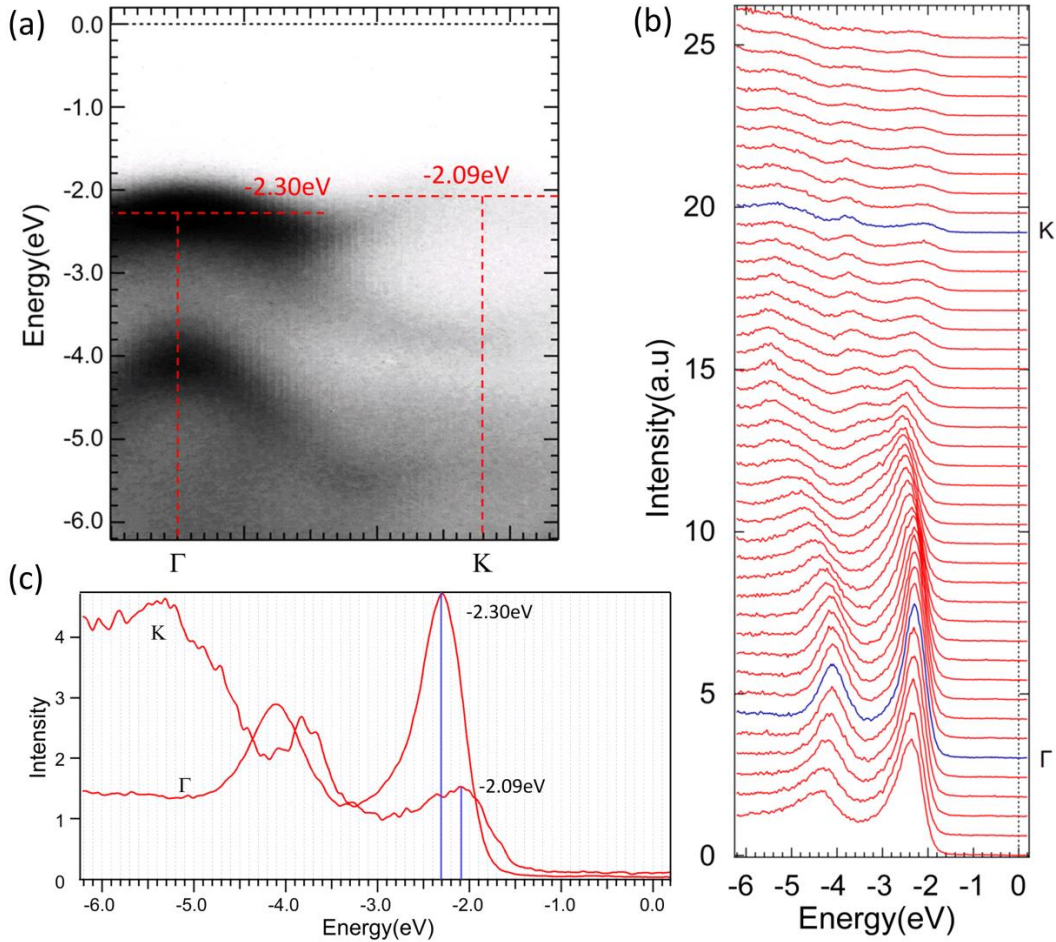

Figure 7: ARPES spectrum and analysis curves of Sample #B. (a) Room temperature ARPES spectrum; (b) Energy distribution curves along the Γ-K direction; (c) The two energy distribution curves obtained at Γ and K points.

**Conclusion**

In summary, a filament-type incandescent source of Molybdenum was invented for the growth of large-area single-crystalline MoSe$_2$ thin films on various substrates, which enjoys several advantages over existing conventional approaches. Detailed Raman spectroscopic studies revealed a new Raman characteristic peak of MoSe$_2$, which is located at 1591 cm$^{-1}$. It was found that the growth on sapphire substrate provides the best smoothness of the resulting thin films with atomic flatness. More importantly, the growth of large-area single monolayer and single crystalline MoSe$_2$ was demonstrated to be achievable using the invented technique with consistent evidences provided by the results from studies using Raman spectroscopy, photoluminescence, RHEED and ARPES. This invented growth technique could be extended to fabricate other 2D TMDCs materials containing a low-vapor-pressure element, such as NbSe$_2$ and WSe$_2$, and also will pave the way for realizing many promising large-scale applications based on these materials.

**Acknowledgement:**

We gratefully acknowledge the use of the facilities in the Materials Characterization and Preparation Facility (MCPF) at the Hong Kong University of Science and Technology (HKUST). The work described here was substantially supported by internal resources from the Department of Physics of HKUST. Work at SUSTech was supported by the National Natural Science Foundation of China (NSFC) under Grant No. 11504159, NSFC Guangdong under Grant No. 2016A030313650, and the Shenzhen Science and Technology Innovations Committee under Projects No. JCY20150630145302240, JCYJ2016053119024691 & KQCX20140522151322951.

# Supplementary Information

## Large-area epitaxial growth of MoSe2 via an incandescent molybdenum source

Man-Kit Cheng[1,4], Jing Liang[1,4], Ying-Hoi Lai[1], Liang-Xi Pang[1], Yi Liu[1], Junying Shen[1], Jianqiang Hou[1], Qing Lin He[1], Bochao Xu[2], Junshu Chen[2], Gan Wang[2], Chang Liu[2], Rolf Lortz[1] & Iam-Keong Sou[1,3]

[1] Department of Physics, the Hong Kong University of Science and Technology, Clear Water Bay, Hong Kong SAR, China

[2] Department of Physics, Southern University of Science and Technology, No. 1088, Xueyuan Rd., Xili, Nanshan District, Shenzhen, Guangdong, China

[3] William Mong Institute of Nano Science and Technology, the Hong Kong University of Science and Technology, Clear Water Bay, Hong Kong SAR, China

[4] Contributed equally to this work

E-mail: phiksou@ust.hk

**Supplementary Information:**

Supplementary Figures 1-3

Supplementary References 1

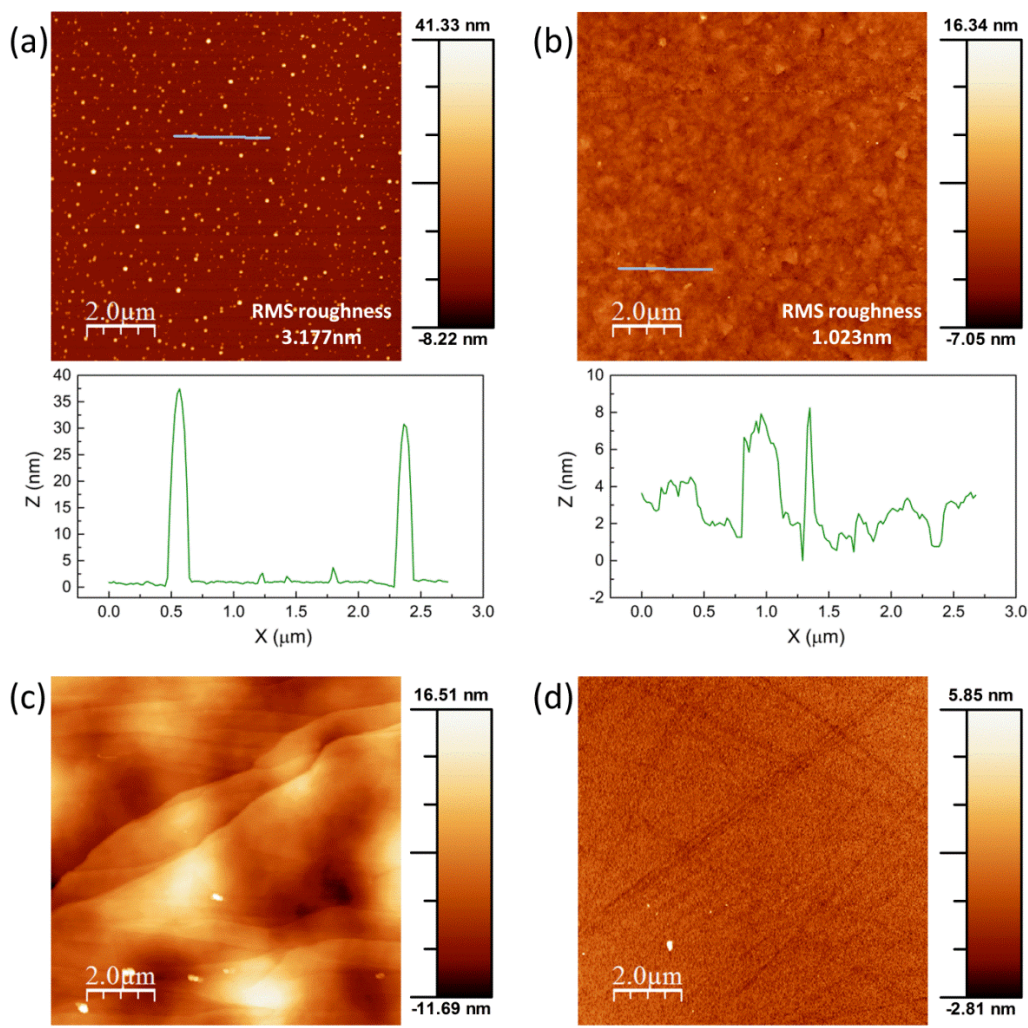

**Supplementary Figure S1**

Typical AFM images with a scan area of 10 × 10 µm$^2$ together with line profiles for MoSe$_2$ layers grown on (a) bare GaAs(111)B, (b) Bi$_2$Te$_3$/GaAs(111)B. The surface of the MoSe$_2$ thin film grown on a bare GaAs(111) B substrate consists of high-density islands with heights around a few nanometers and widths around 160nm. The surface morphology of the MoSe$_2$ layer grown on the Bi$_2$Te$_3$/GaAs(111)B is much improved but still suffer from a surface roughness around 1-2 nm. (c) AFM image of a pure HOPG substrate shows its terrace-like surface. (d) AFM image of a pure sapphire substrate, in which one can see some scratched lines resulted from the manufacturing process.

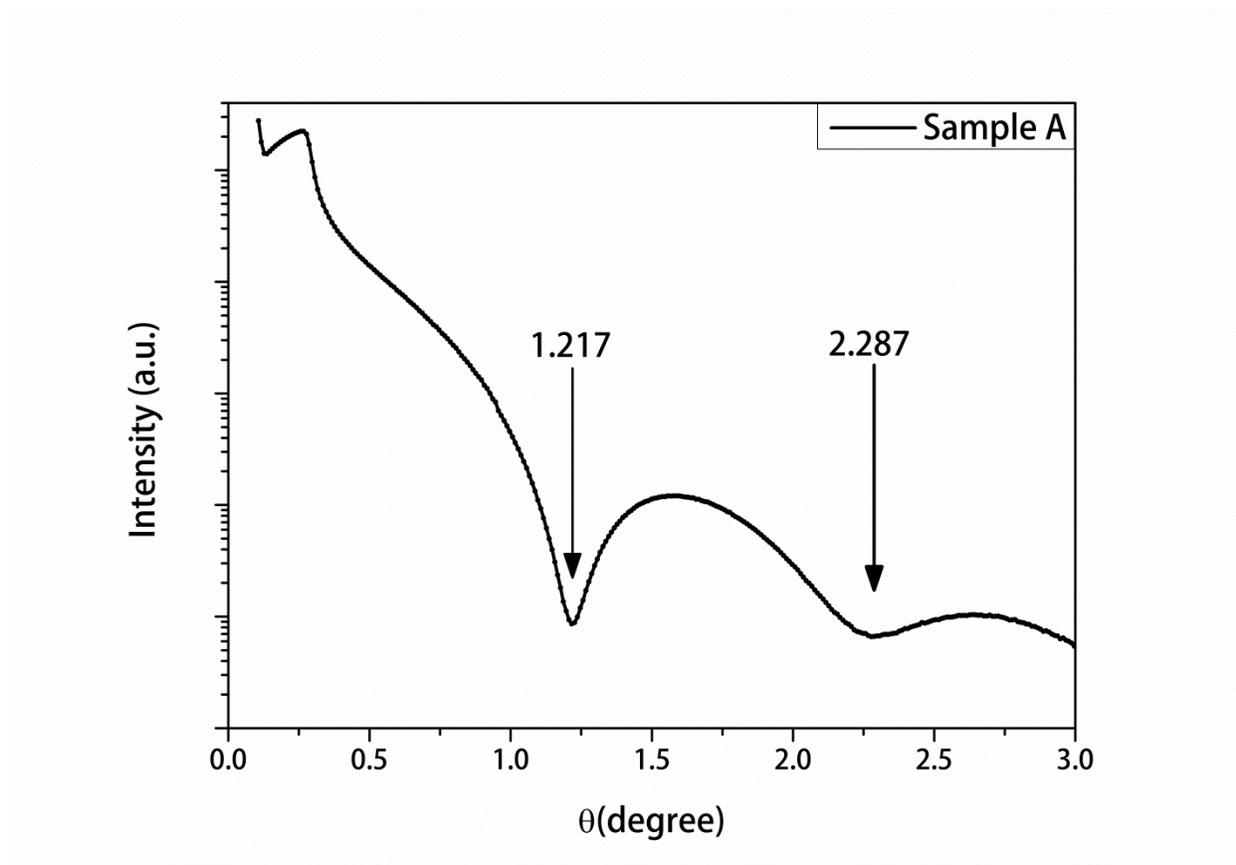

**Supplementary Figure S2**

X-ray reflectivity spectrum of Sample #A measured using the PANalytic Emptrean X-ray reflectometer with $\lambda_{Cu-K\alpha} = 1.5418740\text{Å}$ that housed in The Materials Fabrication and Preparation Facilities of The Hong Kong University of Science and Technology. The angular values of the observed two dips in the spectrum are determined to be $\theta m = 1.217°$ and $\theta m+1 = 2.287°$. The thickness of the MoSe$_2$ layer of this sample was determined using the well-known formula $d \approx \frac{\lambda}{2} \frac{1}{\theta_{m+1} - \theta_m}$ to be ~ 6.4 ML of MoSe$_2$. Taking the uncertainly of the measurements into account, d is estimated to be 6 -7 ML.

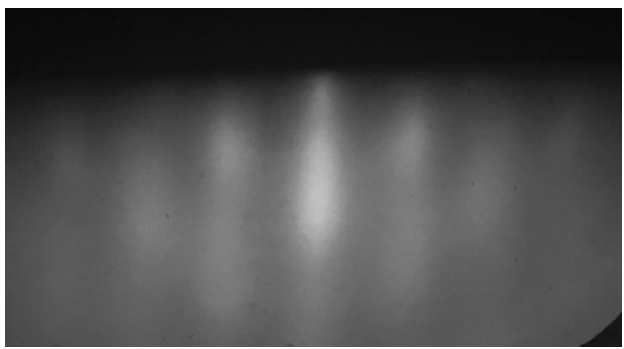

**Supplementary Figure S3**

Observed RHEED (Reflection high-energy electron diffraction) pattern of Sample #B (a single mono-layer of MoSe$_2$) after the Se capping layer was thermally removed in the MBE system housed in the Department of Physics of The Southern University of Science and Technology. In taking this pattern, the e-beam was along the $[11\bar{2}0]$ direction of the MoSe$_2$ lattice. The observed pattern well matches with the corresponding pattern reported by Aretouli, K. E. *et al* [1].

**Supplementary Reference**